\documentclass[10pt]{article}
\usepackage[utf8]{inputenc}
\usepackage{graphicx}
\usepackage{amsmath}
\usepackage{amsfonts}
\usepackage{amssymb}
\usepackage{physics, siunitx}
\usepackage{setspace}
\usepackage{authblk}
\usepackage{caption}
\usepackage{float}
\usepackage{cite}
\usepackage[left=2.0cm,right=2.0cm,top=2cm,bottom=2cm]{geometry}
\usepackage{color}
\usepackage{multicol}
\usepackage{subcaption}
\usepackage{color,soul}
\usepackage{xcolor}

\title{\textbf{Geometric Manifold Statistics of Turbulence-Impacted Beam Propagation and Compensation in Optical Communication}}

\author[1,*]{Shouvik Sadhukhan}
\author[2]{C. S. Narayanamurthy}

\affil[1, 2]{\small{Applied and Adaptive Optics Laboratory, Department of Physics, Indian Institute of Space Science and Technology (IIST), P.O: Valiamala, Trivandrum - 695547, State: Kerala, India}}
\affil[1]{\small{Email: shouvikphysics1996@gmail.com}}
\affil[2]{\small{Email: naamu.s@gmail.com}}
\affil[*]{\small{Corresponding Author Email: shouvikphysics1996@gmail.com}}

\begin{document}
\maketitle

\begin{abstract}
This study develops a manifold-based statistical framework for analyzing turbulence-affected optical beam propagation and compensation in free-space communication systems. The spatial intensity distributions, distorted by dynamic turbulence, are represented using Gaussian Mixture Models (GMMs), whose probability landscapes are refined via Kernel Density Estimation (KDE) applied to pixel-level intensity data across temporal frames. The temporal evolution of turbulence is quantified by monitoring the variation of the unnormalized volumetric integrals under the GMM surfaces, providing a continuous measure of power redistribution within the optical field. Experimental investigations were carried out under four propagation conditions: turbulence-free reference, turbulence only, turbulence with a single Poly(methyl methacrylate) (PMMA) compensator, and turbulence with dual PMMA compensators. To assess statistical dissimilarity, Affine-Invariant Riemannian Metric (AIRM) distances were computed between the covariance representations of successive frames, capturing the geometric evolution of beam topology in the space of Symmetric Positive Definite (SPD) matrices. Complementary topological distance analyses, performed with respect to the initial frame of each set, revealed distinct signatures of turbulence-induced aberrations and their progressive mitigation through dielectric compensation. The observed reduction in inter-frame distance metrics confirms the ability of coupled dipole interactions within PMMA slabs to partially restore statistical coherence in turbulence-distorted optical beams.\\

\end{abstract}

\section{Introduction}
Free-space optical (FSO) communication systems face significant challenges from atmospheric turbulence, which introduces random fluctuations in intensity, phase, and polarization \cite{2,3,4,5}. These distortions manifest as beam wander, scintillation, and coherence loss, degrading link reliability. Traditional mitigation approaches include adaptive optics requiring complex wavefront sensing and active correction \cite{6,7}, diversity techniques combining multiple beams or receivers \cite{8,9}, and passive compensation through engineered optical elements \cite{10,11}.

Recent advances in information geometry and statistical manifold theory \cite{68,72,73} provide powerful tools for characterizing probability distributions. The Fisher-Rao metric establishes an intrinsic Riemannian structure on the manifold of probability densities \cite{88}, enabling coordinate-invariant quantification of statistical dissimilarity. This formalism has been extended to Gaussian Mixture Models (GMMs) through embeddings into the manifold of Symmetric Positive Definite (SPD) matrices \cite{1,79,80}, facilitating geometric distance measurements via the Affine-Invariant Riemannian Metric (AIRM).

Existing turbulence characterization methods rely primarily on temporal statistics such as scintillation index and correlation functions \cite{5,8}. While informative, these metrics lack geometric interpretation and do not fully capture the topological evolution of beam intensity distributions. Moreover, passive compensation mechanisms through dielectric media remain underexplored compared to active systems. This work addresses these gaps by: (1) developing a unified statistical-geometric framework combining GMM/KDE modeling with manifold-based distance metrics, (2) investigating PMMA-based passive turbulence compensation through controlled experiments, and (3) establishing connections between microscopic dipole dynamics and macroscopic statistical observables.

Our approach integrates two complementary perspectives. From the statistical side, we employ Fisher-Rao geometry and AIRM distances to quantify distributional changes on curved manifolds \cite{1,68}. These tools provide coordinate-invariant measures of how turbulence-induced perturbations evolve the beam's statistical structure.

From the physical side, light propagation through dielectric media induces oscillating dipoles in the material's electron clouds. When electromagnetic waves traverse transparent media like PMMA, the electric field perturbs molecular electron distributions, creating induced dipole moments. In dense media, these dipoles do not oscillate independently—they couple through electromagnetic interactions described by dyadic Green's functions \cite{17,21,22}. This coupling leads to collective oscillation modes that can resist rapid field redistributions caused by turbulence. The physical mechanism underlying PMMA compensation relies on three effects: (1) \textit{Dipole-dipole coupling} creates synchronized collective modes resistant to individual perturbations \cite{27,31,32}; (2) \textit{Gradient forces} from spatial intensity variations induce position-dependent forces that counteract beam spreading \cite{33,37,38}; (3) \textit{Inertial effects} in synchronized dipole systems oppose rapid field redistribution, acting as temporal low-pass filters \cite{62,91}. Unlike active adaptive optics that correct wavefront distortions through real-time feedback, PMMA compensation operates passively by stabilizing the beam's statistical distribution through material-mediated dipole dynamics. This provides an inherently stable, power-independent approach complementary to active techniques.

The remainder of this paper is organized as follows. Section 2 presents the simplified theoretical framework with detailed derivations in the Appendix and supplementary file. Section 3 describes the statistical background and physical interpretation of distance metrics. Section 4 contains brief of the predictions that has been validated in this work. Section 5 details the experimental setup with complete parameter specifications. Section 6 analyzes results and validates theoretical predictions. Section 7 concludes with discussion of implications and future directions.

\section{Theoretical Framework}\label{2}
This section presents the core theoretical concepts underlying PMMA-based turbulence compensation. Detailed mathematical derivations are provided in Appendix A.

\subsection{Lorentz Dipole Oscillator Model}
When an electromagnetic wave propagates through a transparent medium, its electric field $\mathbf{E}_{\text{ext}}(\mathbf{r}_i, t)$ perturbs the electron clouds of constituent molecules, inducing oscillating dipoles. The displacement $\mathbf{r}_i$ of the $i$-th electron from equilibrium is governed by:

\begin{equation}
    m \ddot{\mathbf{r}}_i + m \gamma \dot{\mathbf{r}}_i + m \omega_0^2 \mathbf{r}_i = -e \mathbf{E}_{\text{ext}}(\mathbf{r}_i, t)
    \label{eq:lorentz_basic}
\end{equation}

where $m$ is the effective electron mass, $\gamma$ is the damping constant, $\omega_0$ is the resonant frequency, and $e$ is the electron charge. For complex molecular structures like PMMA, nonlinear restoring forces become important:

\begin{equation}
    m \ddot{\mathbf{r}}_i + m \gamma \dot{\mathbf{r}}_i + m \omega_0^2 \mathbf{r}_i + \beta_i |\mathbf{r}_i| \mathbf{r}_i + \alpha_i |\mathbf{r}_i|^2 \mathbf{r}_i = -e \mathbf{E}_{\text{ext}}(\mathbf{r}_i, t)
    \label{eq:anharmonic}
\end{equation}

where $\beta_i$ and $\alpha_i$ represent second- and third-order nonlinearity coefficients, respectively. In the present context of work the PMMA shows very little higher-order nonlinear susceptibility for wavelength 632.8 nm; hence, we can consider $\beta_i\rightarrow 0$ and $\alpha_i\rightarrow 0$.

\subsection{Dipole-Dipole Coupling and Collective Modes}
In dense media, dipoles interact through electromagnetic coupling mediated by the dyadic Green's function $\mathbf{G}(\mathbf{r}_i, \mathbf{r}_j)$ (see Appendix A for derivation). This coupling introduces additional forces:

\begin{equation}
    \mathbf{F}_{\text{coupling}} = e^2 \mu_0 \omega^2 \sum_{j \ne i} \mathbf{G}(\mathbf{r}_i, \mathbf{r}_j) \cdot \mathbf{r}_j(t)
    \label{eq:coupling_force}
\end{equation}

When spatial field gradients are present, gradient forces arise:

\begin{equation}
    \mathbf{F}_{\text{grad}} = \sum_{j \ne i} [e(\mathbf{r}_j - \mathbf{r}_i)\cdot\nabla]\mathbf{E}_{\text{ext}}(\mathbf{r}_j, t)
    \label{eq:gradient_force}
\end{equation}

After application of gradient force and coupling force the force counterpart of Lorentz oscillation equation should be modified as $-e \mathbf{E}_{\text{ext}}(\mathbf{r}_i, t)\rightarrow-e \mathbf{E}_{\text{ext}}(\mathbf{r}_i, t)+\mathbf{F}_{\text{coupling}}+\mathbf{F}_{\text{grad}}$. The complete dynamics include these coupling terms, which can be diagonalized to reveal collective oscillation modes (Appendix A). These collective modes synchronize dipole oscillations, creating stability against turbulent perturbations. Detailed derivations are given in supplementary materials. The compensation mechanism operates through three synergistic effects:

\begin{itemize}
    \item \textbf{Synchronization:} Coupled dipoles establish collective modes with characteristic timescales $\tau_{\text{sync}} \sim L/c\sqrt{\epsilon_r}$, where $L$ is the PMMA length and $\epsilon_r \approx 2.2$ is its relative permittivity. For turbulence fluctuations occurring on timescales $\tau_{\text{turb}}$, effective compensation requires $\tau_{\text{sync}} < \tau_{\text{turb}}$.

    \item \textbf{Gradient Stabilization:} Spatial intensity gradients induce forces (Eq. \ref{eq:gradient_force}) opposing beam spreading. This effect scales with PMMA length and refractive index contrast.

    \item \textbf{Inertial Damping:} Synchronized dipole systems exhibit effective inertia that resists rapid field redistribution. The perturbation force during turbulence fluctuations is:

\begin{equation}
    \delta \mathbf{F}_{\text{pert}} = \mathbf{F}'_{\text{inertia}} - \mathbf{F}_{\text{inertia}}
    \label{eq:pert_force}
\end{equation}
\end{itemize}

When synchronization is established, $|\delta \mathbf{F}_{\text{pert}}|$ becomes small, explaining reduced temporal fluctuations. 

\subsection{Susceptibility-Induced Spectral Filtering and Scintillation Reduction}

The dipole-dipole coupling and Lorentz model produce the fluctuating susceptibility through the polarizability solution. Optical beam propagation through dielectric media with microscopic susceptibility fluctuations can be described using the inhomogeneous Helmholtz equation
\begin{equation}
(\nabla^2 + k_0^2)E(\mathbf r)
=
- k_0^2 \chi(\mathbf r)E(\mathbf r),
\end{equation}
where spatial variations in $\chi(\mathbf r)$ act as distributed scattering sources that redistribute spatial frequencies. Under turbulence impacted fluctuations $\chi(\mathbf r)$ must become $\chi(\mathbf r)+\delta\chi(\mathbf r)$. This $\delta\chi(\mathbf r)$ factor produces turbulence impacted scattering inside the medium of propagation that again produces $\gamma(k)$. Detail derivation has been attached into supplementary file.

Under weak scattering and forward propagation, cumulative interactions lead to a multiplicative transfer function in the spatial-frequency domain,
\begin{equation}
H(k)=t\,\exp\!\left[i\frac{k_0 \bar{\chi} L}{2}\right]\exp[-\gamma(k)L],
\end{equation}
where $\gamma(k)\propto S_\chi(k)$ represents scattering loss and $t=e^{-\alpha L}$ accounts for absorption and interface losses.

Because $\gamma(k)$ increases with spatial frequency, the dielectric medium suppresses high-frequency fluctuations responsible for speckle and scintillation. Usually we can take $\mathrm{Var}[I] =
4|E_0|^2
\int \Phi_{\delta E}(\mathbf k)\, d^2k$. The resulting reduction in intensity variance is governed by
\begin{equation}
\mathrm{Var_{out}}[I] \propto \int e^{-2\gamma(k)L}\,\Phi_{\delta E}(k)\, dk,
\end{equation}
demonstrating that susceptibility fluctuations act as a spatial low-pass filter. Increasing propagation length enhances fluctuation suppression but reduces transmitted power, implying an optimal length for turbulence mitigation. In the context of present work, temporal fluctuation or scintillation $\frac{\sigma_{out}^2}{\sigma_{in}^2}\propto\frac{\mathrm{Var_{out}}[I]}{\mathrm{Var_{in}[I]}}\propto \exp[-\gamma(k)L]$, thus, the statistical distance of the turbulence impacted data frames w.r.t., corresponding first frame are proportional to $\exp[-\gamma(k)L]$ or $\exp(-L/L_0)$ with characteristic length $L_0$ which came from susceptibility fluctuation of the system. Detailed derivations are given in supplementary materials. 

\section{Statistical Background}\label{3}
This section describes the statistical framework used to quantify turbulence effects and compensation. We explain the physical meaning of each metric in the context of optical beam analysis. Detailed derivations are given in supplementary materials. The statistical framework described below enables quantification of how dipole synchronization manifests in macroscopic observables. While Section 2 described microscopic coupling mechanisms, this section establishes the geometric-statistical language for characterizing their collective effects on beam intensity distributions.

\subsection{Gaussian Mixture Models and Kernel Density Estimation}
Beam intensity distributions are modeled as Gaussian Mixture Models:

\begin{equation}
    p(x; \theta) = \sum_{k=1}^K \pi_k \mathcal{N}(x \mid \mu_k, \Sigma_k)
    \label{eq:gmm}
\end{equation}

where $\theta = \{(\pi_k, \mu_k, \Sigma_k)\}_{k=1}^K$ parameterizes component weights, means, and covariances. For different data set we $q(x; \theta)$ has been chosen in place of $p(x; \theta)$. For $K=5$ components, this captures multi-modal structures characteristic of turbulence-affected beams. The Kernel Density Estimation provides non-parametric smoothing:

\begin{equation}
    \hat{f}_H(\mathbf{x}) = \frac{1}{n (2\pi) |H|^{1/2}} \sum_{i=1}^{n} \exp\left(-\frac{1}{2} (\mathbf{x} - \mathbf{x}_i)^\top H^{-1} (\mathbf{x} - \mathbf{x}_i)\right)
    \label{eq:kde}
\end{equation}

where $H$ is the bandwidth matrix selected via cross-validation.

\textit{Physical Interpretation:} GMM components represent distinct intensity regions (central peak, side lobes, scattered light), while KDE smooths pixel-level noise to reveal underlying probability structure.

\subsection{Statistical Distance Metrics}
We employ multiple distance measures, each capturing different aspects of distributional similarity:

\begin{itemize}
    \item \textbf{Kullback-Leibler Divergence:}
\begin{equation}
    D_{KL}(p \| q) = \int p(x) \log \frac{p(x)}{q(x)} dx
    \label{eq:kl}
\end{equation}
\textit{Physical Meaning:} Information-theoretic measure of how turbulence distorts the probability structure. Large $D_{KL}$ indicates significant redistribution of optical power away from the reference distribution.

    \item \textbf{Jensen-Shannon Divergence:}
\begin{equation}
    D_{JS}(p \| q) = \frac{1}{2} D_{KL}(p \| m) + \frac{1}{2} D_{KL}(q \| m), \quad m = \frac{p+q}{2}
    \label{eq:js}
\end{equation}

\textit{Physical Meaning:} Symmetric, bounded version of KL divergence. $\sqrt{D_{JS}}$ forms a true metric quantifying statistical dissimilarity robust to outliers.

    \item \textbf{Bhattacharyya Distance:}
\begin{equation}
    D_B(p, q) = -\ln \int \sqrt{p(x) q(x)} dx
    \label{eq:bhatt}
\end{equation}

\textit{Physical Meaning:} Measures probabilistic overlap. In optical context, $\exp(-D_B)$ represents the fidelity of intensity distribution preservation under turbulence.
\end{itemize}

There are four sets of results with 200 frames in each of them. Each frame is smoothed using KDE density function and the volume under the smoothed images after normalization are found to get power distributions. The power distributions for those four sets of data are used to find the discussed divergences.

\subsection{Affine-Invariant Riemannian Metric (AIRM)}
GMM parameters are embedded into the manifold of SPD matrices $\text{SPD}_{K(n+1)}(\mathbb{R})$ via:

\begin{equation}
    f(\theta) = \begin{pmatrix} A & X \\ X^T & B \end{pmatrix}
    \label{eq:spd_embedding}
\end{equation}

where $A = \text{diag}(\Sigma_1 + \pi_1 \mu_1 \mu_1^T, \ldots, \Sigma_K + \pi_K \mu_K \mu_K^T)$, $X = \text{diag}(\pi_1 \mu_1, \ldots, \pi_K \mu_K)$, and $B = \text{diag}(\pi_1, \ldots, \pi_K)$.

The AIRM distance between embedded matrices $P, Q$ is:

\begin{equation}
    d_{\text{AIRM}}(P, Q) = \left\| \log(P^{-1/2} Q P^{-1/2}) \right\|_F
    \label{eq:airm}
\end{equation}

\textit{Physical Meaning:} Geodesic distance on the SPD manifold capturing geometric evolution of beam topology. AIRM respects the intrinsic curvature of the statistical manifold, providing coordinate-invariant quantification of how turbulence transforms the beam's multi-dimensional intensity structure. Unlike Euclidean metrics that treat all parameter changes equally, AIRM accounts for the non-flat geometry where certain distributional changes are "closer" than others in information-theoretic sense.

In turbulence analysis, AIRM distance fluctuations reveal temporal stability: small frame-to-frame AIRM distances indicate stable statistical structure despite ongoing turbulence, while large fluctuations signal rapid topological changes.

\subsection{Power Metrics}
Unnormalized GMM volume quantifies total optical power:

\begin{equation}
    V_i = \sum_{k=1}^{K}\pi_k 2\pi\sqrt{|\Sigma_k|}
    \label{eq:gmm_power}
\end{equation}

\textit{Physical Meaning:} Temporal fluctuations in $V_i$ reflect turbulence-induced scintillation. Reduced fluctuation range with PMMA compensation indicates stabilized power distribution. The power is fluctuating due to fluctuation of best fitting of GMM on each frame of those four sets of data.

\section{Predictions to validate}
The theoretical framework predicts quantifiable relationships between system parameters and compensation effectiveness. These predictions are validated experimentally in Section 6.

\begin{itemize}
    \item \textbf{Prediction 1 (Scaling with Length):} Statistical distance fluctuations should decrease as $\Delta d_{\text{AIRM}} \propto \exp(-L/L_0)$ where $L$ is total PMMA length and $L_0$ is a characteristic synchronization length. The exponential relation can be found from the equation 8 in section 2.

    \item \textbf{Prediction 2 (Turbulence Strength Dependence).}
The effectiveness of dielectric compensation depends on the turbulence coherence length $r_0$ relative to a material-dependent critical scale $r_{\mathrm{crit}}$. Turbulence introduces high spatial-frequency fluctuations with characteristic scale $\sim r_0$, producing fine speckle and rapid decorrelation. Propagation through PMMA acts as a spatial low-pass filter with transfer function $|H(\kappa)|^2=\exp[-2\gamma(\kappa)L]$, which suppresses high spatial frequencies associated with small-scale fluctuations. This defines a cutoff spatial scale
\[
r_{\mathrm{crit}} \sim \frac{1}{\kappa_c},
\]
below which fluctuations are strongly attenuated. Consequently,
\[
r_0 > r_{\mathrm{crit}} \Rightarrow \text{effective compensation}, \qquad
r_0 < r_{\mathrm{crit}} \Rightarrow \text{reduced effectiveness}.
\]
Experimentally, $r_{\mathrm{crit}}$ can be estimated from the minimum stabilized speckle scale after compensation, representing the smallest spatial structure the dielectric medium can suppress.

    \item \textbf{Prediction 3 (Temporal Stability):} Frame-to-frame AIRM distance should scale with the ratio $\tau_{\text{sync}}/\tau_{\text{turb}}$.
\end{itemize}

\section{Experimental Setup}\label{4}

\subsection{Optical System Configuration}
\begin{figure}[H]
\centering
\includegraphics[width=1.0\textwidth]{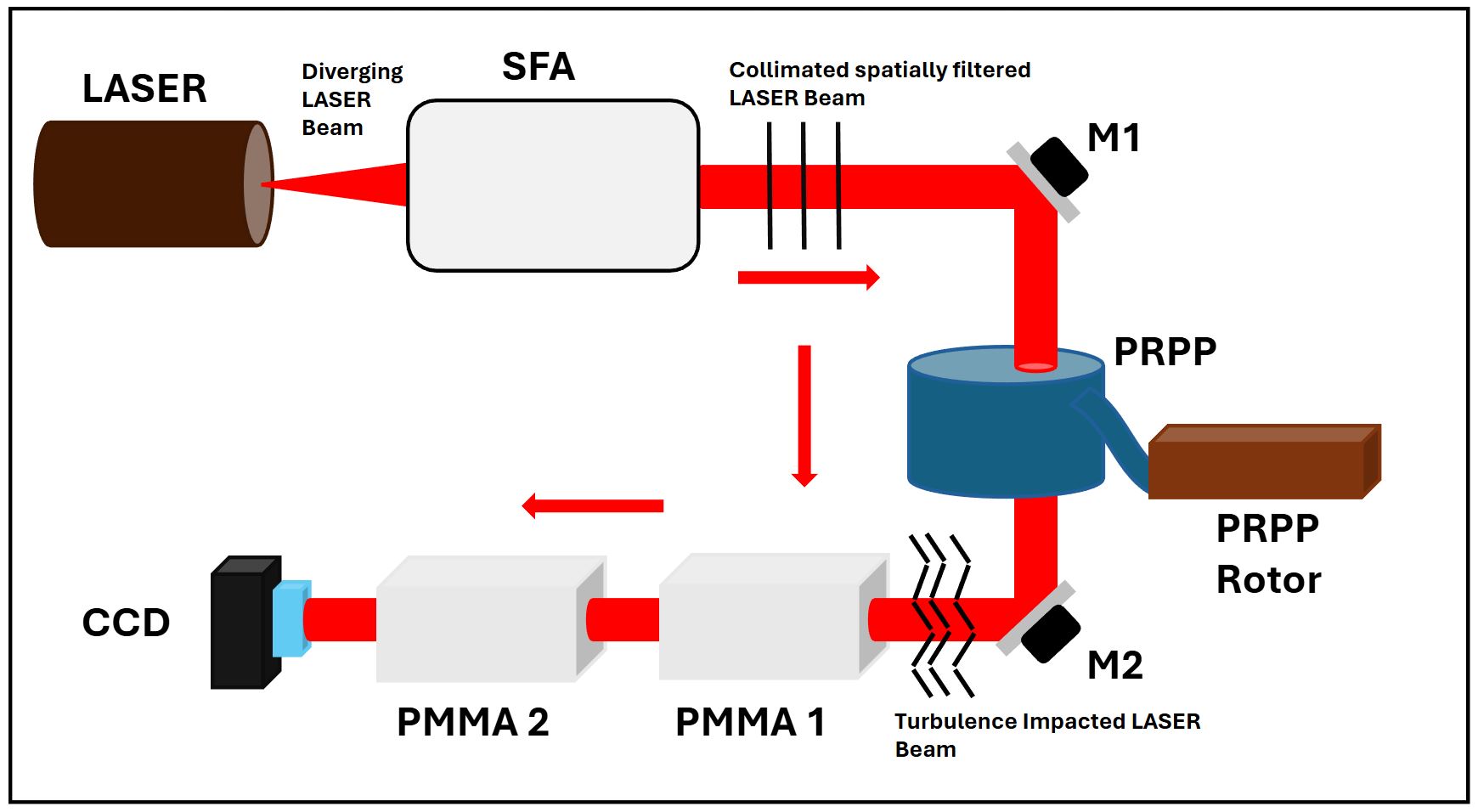}
\caption{Experimental set up for common path beam propagation through turbulence and its compensation methodology. Experimental setup showing laser source (L), spatial filter assembly (SFA), mirrors (M1, M2), pseudo-random phase plate (PRPP), PMMA rods, and CCD camera. Total propagation distance: 1.5 m.}
\label{P0}
\end{figure}

Figure \ref{P0} shows the experimental setup. A continuous-wave He-Ne laser (wavelength $\lambda = 632.8$ nm, output power $P = 5$ mW, beam diameter $d_0 = 6$ mm) was spatially filtered using a 10$\times$ microscope objective and 25 $\mu$m pinhole to generate a clean Gaussian beam. Two mirrors (M1, M2) provided alignment control. 

\subsection{Turbulence Simulation}
Atmospheric turbulence was simulated using a Pseudo-Random Phase Plate (PRPP, Thorlabs EDU-RPP1) consisting of five layers: two BK7 glass windows enclosing a central acrylic layer with imprinted Kolmogorov-type phase profile, stabilized by near-index-matching polymer layers. The plate generates aberrated wavefronts with adjustable Fried coherence lengths $r_0 \approx 0.6$ mm where $16$-$32$ samples distributed over 4096 phase points. The PRPP was mounted on a motorized rotation stage (rotation speed: 1 rpm) to simulate temporal turbulence dynamics.

\textit{Turbulence Scale Equivalence:} In laboratory turbulence simulations such as PRPP experiments, a Fried coherence length as small as $r_0 = 0.6\,\text{mm}$ does not represent natural atmospheric conditions but rather an artificially scaled turbulence strength designed to emulate long-distance atmospheric propagation within a short experimental path. The Fried parameter is given by
\[
r_0 = \left[0.423\, k_0^2 \int_0^L C_n^2(z)\, dz \right]^{-3/5},
\]
and for uniform turbulence simplifies to $r_0 = [0.423\, k_0^2 C_n^2 L]^{-3/5}$. Inverting this expression for $r_0 = 0.6\,\text{mm}$ at $\lambda = 632.8\,\text{nm}$ and $L = 1.5\,\text{cm}$ yields $C_n^2 \approx 3.35\times10^{-7}\,\text{m}^{-2/3}$, which is many orders of magnitude stronger than typical atmospheric values ($10^{-17}$–$10^{-13}\,\text{m}^{-2/3}$). This enhanced $C_n^2$ compensates for the short laboratory propagation length, effectively compressing kilometer-scale atmospheric turbulence into a meter-scale setup. Consequently, the small $r_0$ value reflects a strong turbulence regime used for controlled experimental emulation rather than natural atmospheric coherence conditions. The present scenario provides an equivalent atmospheric turbulence with a minimum 560 km propagation with the same Fried coherence length as PRPP. Detailed derivation has been provided in the supplementary file.

\subsection{PMMA Compensation Elements}
PMMA rods (polymethyl methacrylate) served as passive compensation elements. Key properties:

\begin{itemize}
\item Refractive index: $n_{\text{PMMA}} = 1.492$ at $\lambda = 632.8$ nm
\item Length: $L = 150$ mm per rod
\item Width: $W = 25$ mm
\item Transmission: $T > 92\%$ at $\lambda = 632.8$ nm
\item Surface quality: 40-20 scratch-dig
\item Absorption coefficient: $\alpha < 0.01$ cm$^{-1}$
\end{itemize}

PMMA was selected for several reasons: (1) High optical quality with low absorption/scattering in visible range; (2) Strong dipole-dipole coupling due to polar C=O and C-O-C groups in polymer structure; (3) Sufficient length for synchronization effects ($L/\lambda \approx 10^5$); (4) Availability and cost-effectiveness for practical systems.

\subsection{Detection System}
A 8-bit CCD camera (Thorlabs DCC1545M, pixel size: 5.2 $\mu$m, sensor size: 1280$\times$1024 pixels, quantum efficiency $>$ 60\% at 633 nm) recorded beam intensity distributions. Imaging parameters:

\begin{itemize}
\item Exposure time: 10 ms
\item Frame rate: 5 fps
\item Effective pixel resolution: 4.28 $\mu$m (accounting for imaging magnification)
\item Dynamic range: 8-bit (256 levels)
\item Beam sampling: $\sim$600 pixels across FWHM
\end{itemize}

\subsection{Experimental Procedure}
Four experimental sets were conducted, each recording 200 frames (Figure \ref{P1}):

\begin{itemize}
\item \textbf{Set 1:} Raw turbulence (PRPP only, no PMMA)
\item \textbf{Set 2:} Turbulence + 1 PMMA rod ($L = 150$ mm)
\item \textbf{Set 3:} Turbulence + 2 PMMA rods ($L_{\text{total}} = 300$ mm)
\item \textbf{Set 4:} Turbulence-free reference (no PRPP, no PMMA)
\end{itemize}

\textit{Addressing Sample Size Concerns:} While 200 frames per set may appear limited in the present context of laboratory simulated turbulence phase plate. The turbulence phase plate i.e., PRPP (Pseudo Random Phase Plate) has a annular structure which can be rotated using external rotor. Due to the rotation of that phase plate, we found temporal changes in turbulence which have been recorded in CCD in multiple frame. Increase in number of frame increases the time of recording which may bring same turbulence after a complete rotation of the phase plate. Hence, the number of frame has been fixed into 200 such that the repeatation wouldn't be recorded during experiment. Thus, 200 frames may be sufficient in the context of our experiment. Furthermore, the consistency of trends across multiple independent metrics (KL, JS, Bhattacharyya, AIRM) provides robust validation.

\subsection{Data Processing}

Each recorded frame was subjected to a systematic preprocessing and statistical analysis pipeline. First, background subtraction and dark current correction were applied to eliminate detector noise and fixed-pattern offsets. The corrected intensity distribution was then normalized to unit total power according to
\[
I_{\text{norm}}(x,y) = \frac{I(x,y)}{Max(I(x,y))},
\]
ensuring comparability across frames independent of total intensity fluctuations. The normalized data were subsequently modeled using a Gaussian Mixture Model (GMM) with $K=5$ components, where parameters were estimated via the expectation–maximization algorithm. To obtain a smooth nonparametric representation of the intensity probability distribution, Kernel Density Estimation (KDE) was performed with bandwidth matrix $H$ selected through 5-fold cross-validation. Statistical divergence measures were then computed for each frame relative to the Set~4 reference distribution. Finally, the Affine-Invariant Riemannian Metric (AIRM) distance was evaluated by embedding the fitted GMM parameters into the Symmetric Positive Definite (SPD) manifold and computing the corresponding geodesic distance.

\section{Results and Discussion}\label{5}

\subsection{Intensity Distribution Evolution}
\begin{figure}[H]
\centering
\includegraphics[width=1.0\textwidth]{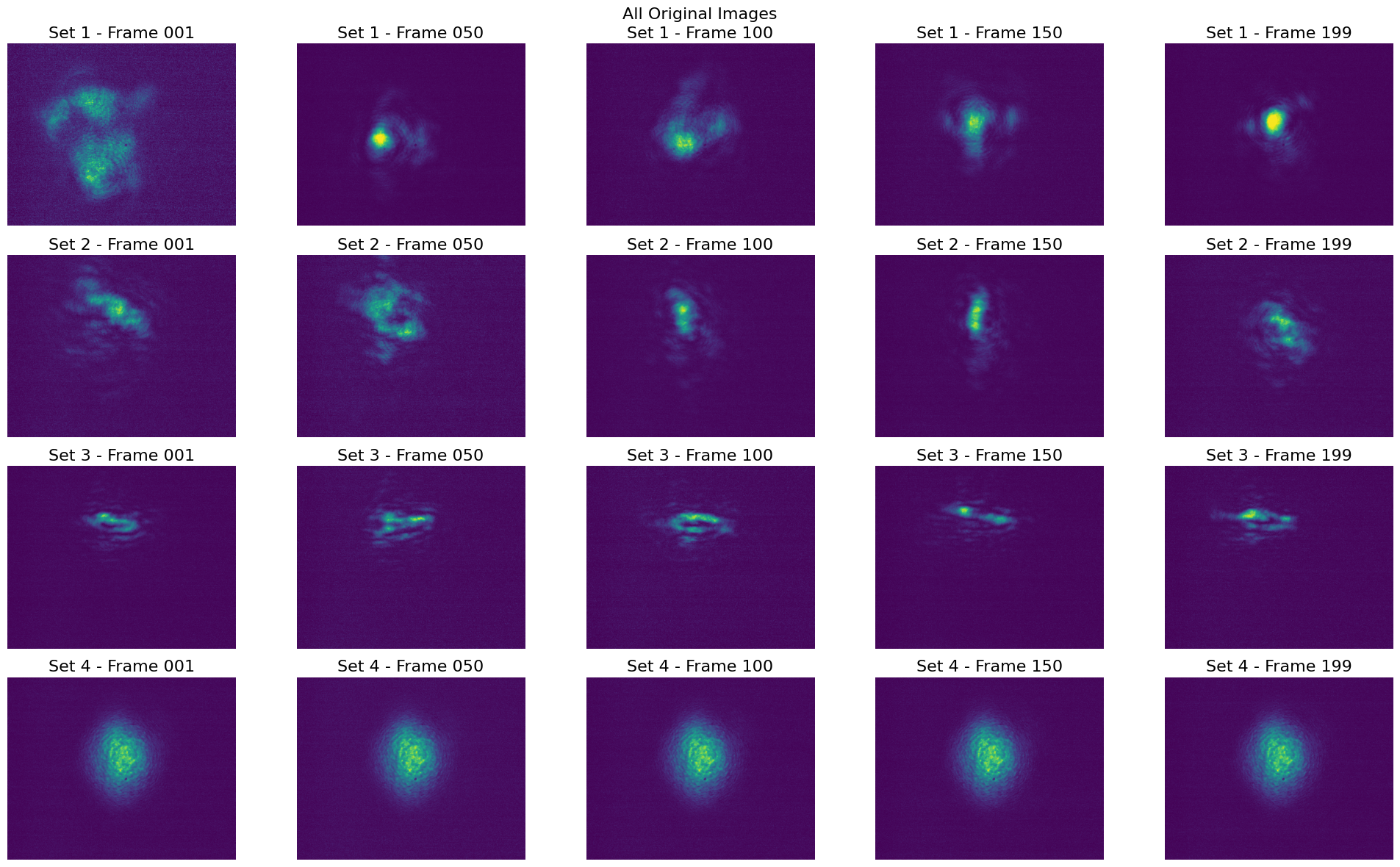}
\caption{Original data for all four sets. Images have been presented after normalization hence, they are unitless.}
\label{fig:ori_all}
\end{figure}

Figures in \ref{fig:ori_all} show representative intensity distributions for each experimental set. The turbulence-free reference (Set 4, Figure \ref{fig:ori_all}) exhibits a stable, symmetric Gaussian profile. Raw turbulence (Set 1, Figure \ref{fig:ori_all}) produces severe distortions with beam spreading, intensity speckles, and asymmetry. Progressive PMMA compensation (Sets 2-3, Figures \ref{fig:ori_all}) restores time-dependent profile fluctuations.

\begin{figure}[H]
\centering
\includegraphics[width=1.0\textwidth]{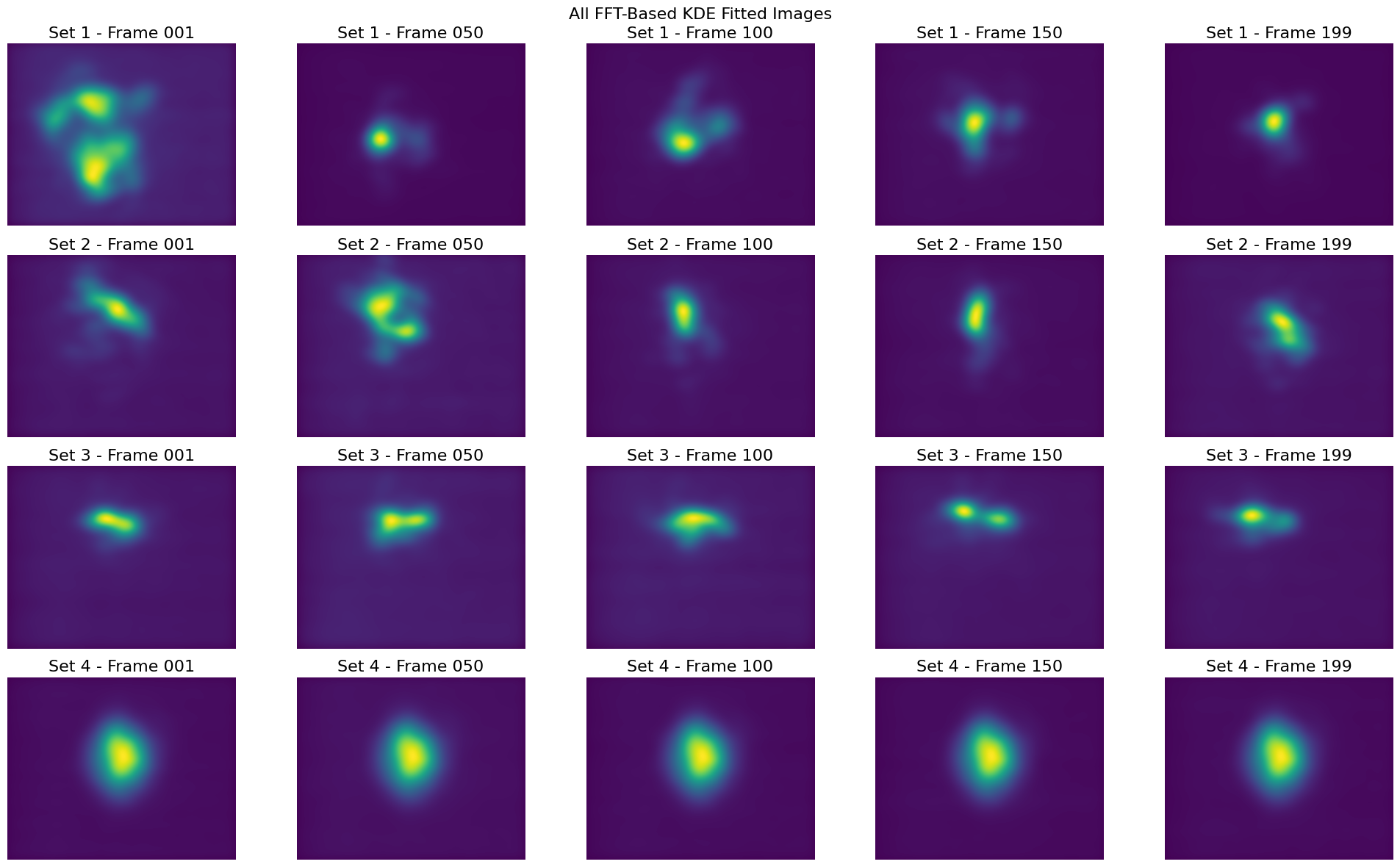}
\caption{KDE Fitted data for all four sets. Fitted images have been presented after normalization hence, they are unitless.}
\label{fig:kde_all}
\end{figure}

Figures in \ref{fig:kde_all} show representative of KDE reconstructions of intensity distributions for each experimental set. The turbulence-free reference (Set 4, Figure \ref{fig:kde_all}) exhibits a stable, symmetric Gaussian profile. Raw turbulence (Set 1, Figure \ref{fig:kde_all}) produces severe distortions with beam spreading, intensity speckles, and asymmetry. Progressive PMMA compensation (Sets 2-3, Figures \ref{fig:kde_all}) restores time-dependent profile fluctuations.

\begin{figure}[H]
\centering
\includegraphics[width=1.0\textwidth]{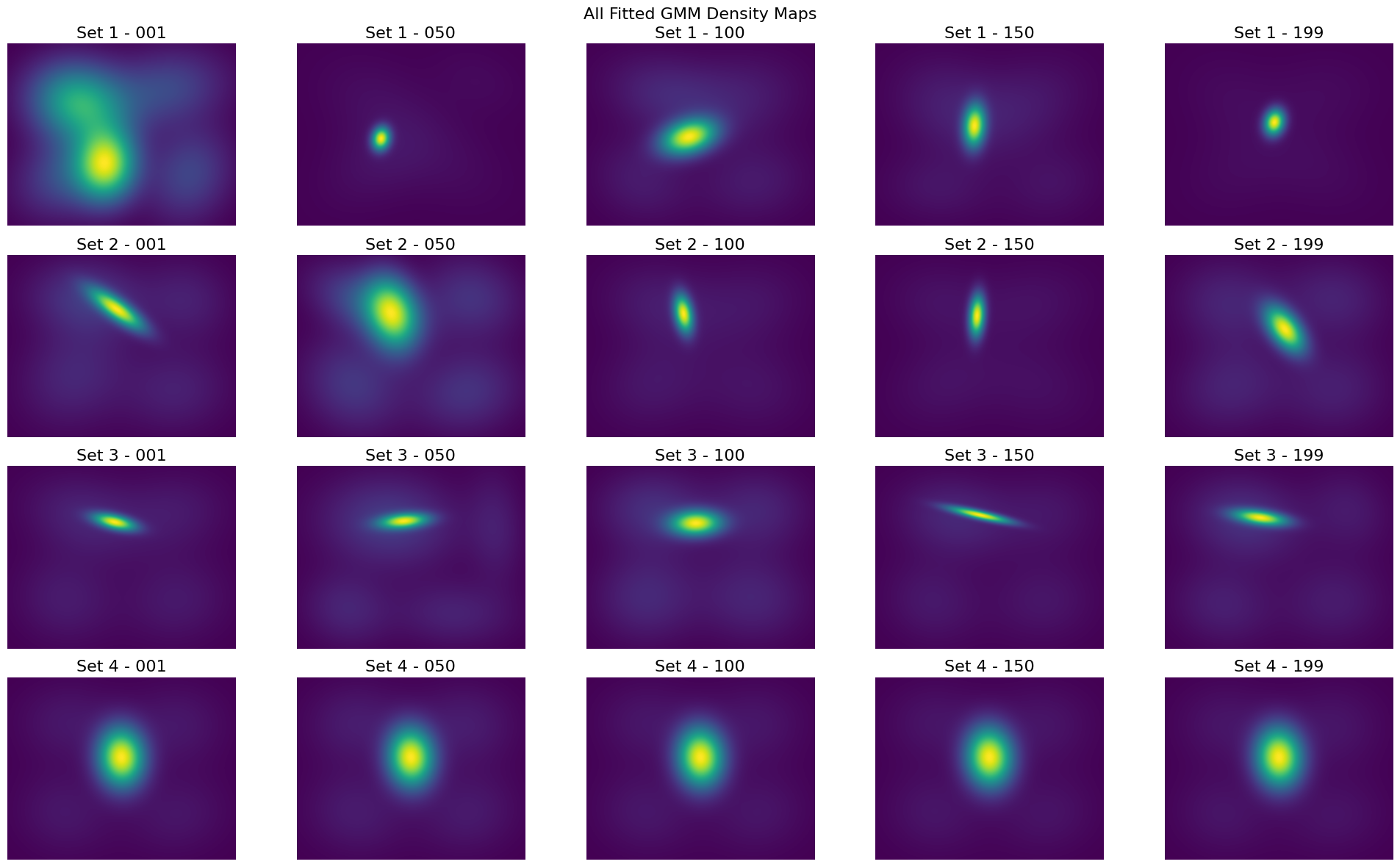}
\caption{GMM Fitted data for all four sets. Fitted images have been presented after normalization hence, they are unitless.}
\label{fig:gmm_all}
\end{figure}

GMM reconstructions (Figures \ref{fig:gmm_all} reveal component-level changes. Set 4 shows tightly clustered components, Set 1 displays widespread component distribution, while Sets 2-3 demonstrate progressive reconcentration—quantitatively capturing coherence restoration.

\subsection{Power Fluctuation Analysis}
\begin{figure}[H]
\centering
\includegraphics[width=0.70\textwidth]{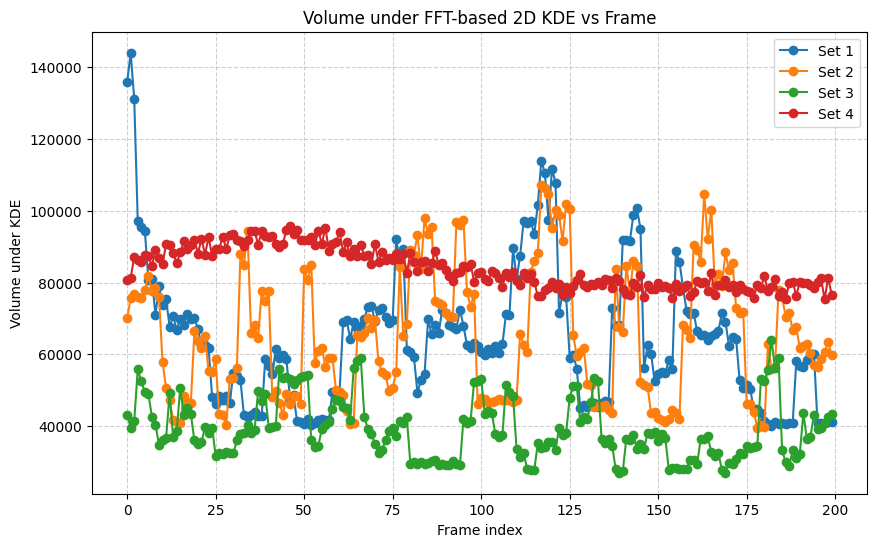}
\caption{Absolute power temporal evolution for all sets. Set 1 (raw turbulence) shows maximum fluctuations; Sets 2-3 demonstrate progressive stabilization with PMMA compensation. Units: arbitrary (normalized to detector counts).}
\label{fig:kde_power_abs}
\end{figure}

Figure \ref{fig:kde_power_abs} shows absolute power under KDE curves across all sets. Set 4 maintains constant power (fluctuation $\approx\pm 6\%$), validating system stability. Sets 1-3 exhibit turbulence-induced fluctuations decreasing systematically with PMMA compensation.

\begin{figure}[H]
\centering
\includegraphics[width=0.70\textwidth]{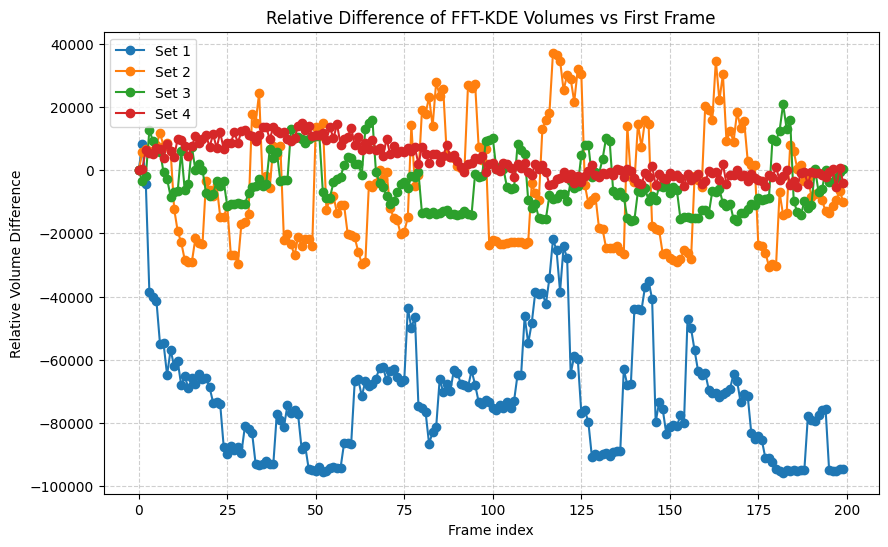}
\caption{Relative power fluctuations. Progressive reduction validates compensation effectiveness. Units: percentage deviation from mean.}
\label{fig:kde_power_rel}
\end{figure}

Relative power fluctuations (Figure \ref{fig:kde_power_rel}) provide normalized comparison. Set 1: $\approx\pm$70\%, Set 2: $\approx\pm$43\%, Set 3: $\approx\pm$15\%, approaching Set 4 stability ($\approx\pm 6\%$). This $\sim$60\% reduction from raw turbulence to dual-PMMA compensation confirms effective scintillation mitigation.

\begin{figure}[H]
\centering
\includegraphics[width=0.70\textwidth]{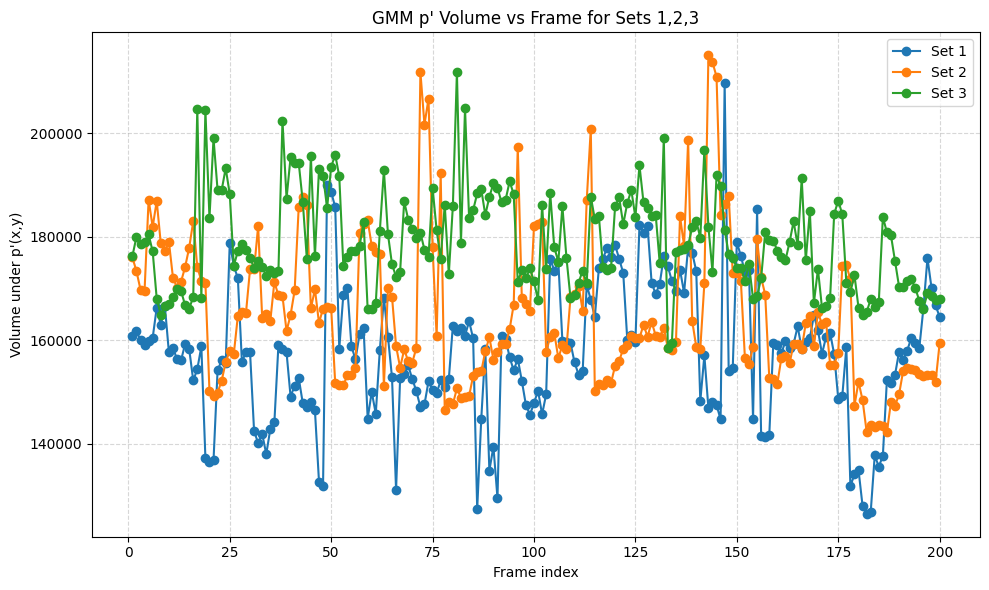}
\caption{GMM-based absolute power for turbulent sets showing consistent reduction in fluctuation amplitude. Units: arbitrary (unnormalized GMM volume).}
\label{fig:gmm_power_abs}
\end{figure}

\begin{figure}[H]
\centering
\includegraphics[width=0.70\textwidth]{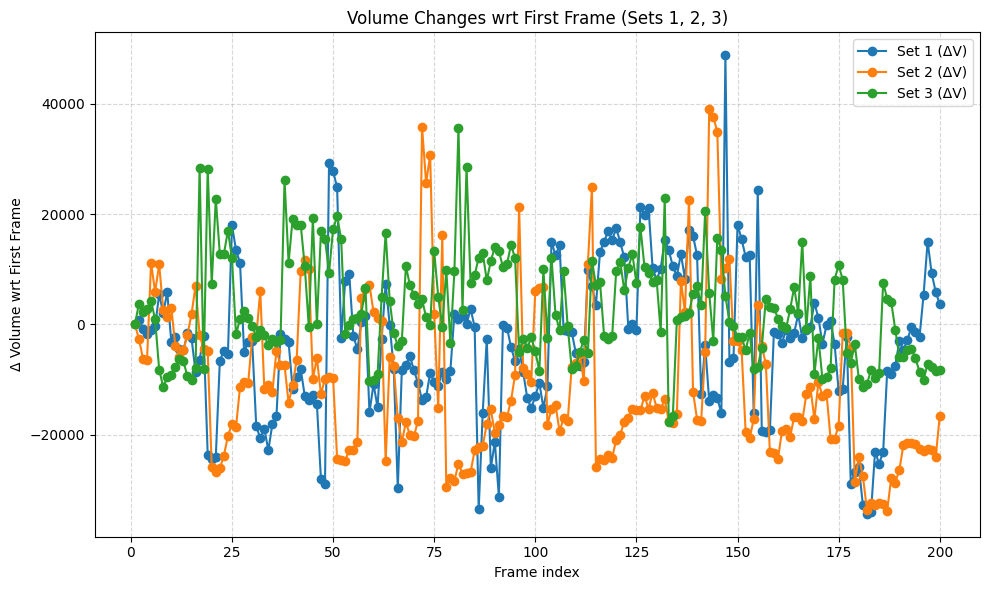}
\caption{GMM-based relative power fluctuations demonstrating improved temporal consistency. Units: percentage deviation from first frame.}
\label{fig:gmm_power_rel}
\end{figure}

GMM-based power analysis (Figures \ref{fig:gmm_power_abs}--\ref{fig:gmm_power_rel}) corroborates KDE results, providing cross-validation through independent statistical methodology.

\subsection{Statistical Distance Metrics}
\begin{figure}[H]
\centering
\includegraphics[width=1.0\textwidth]{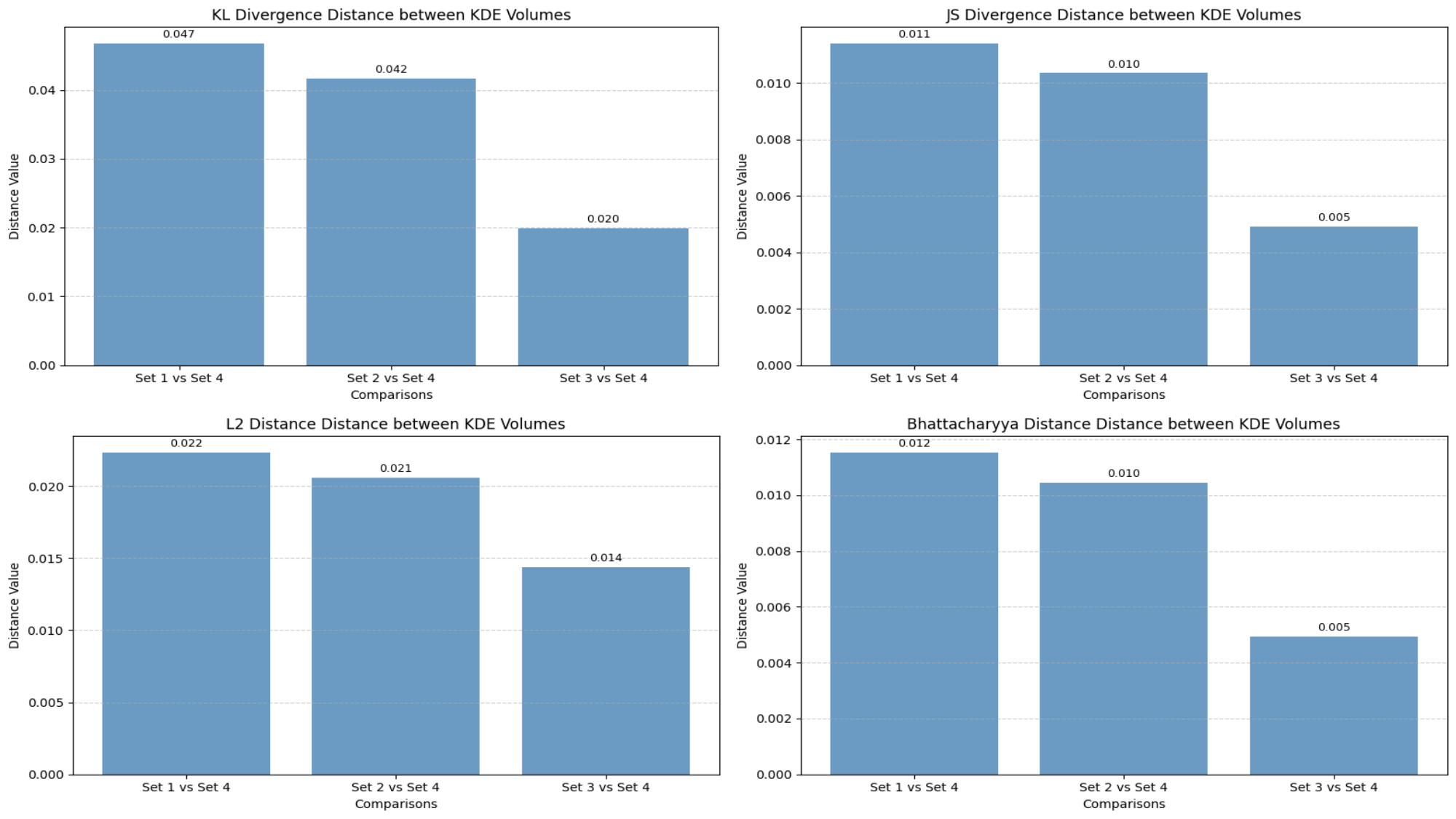}
\caption{Comprehensive statistical distances for KDE distributions. All metrics (KL divergence, JS divergence, L$_2$ distance, Bhattacharyya distance) consistently show decreasing trends with PMMA compensation. Units vary by metric (see legend).}
\label{fig:kde_dist_all}
\end{figure}

Figure \ref{fig:kde_dist_all} presents all four statistical distances (KL, JS, L$_2$, Bhattacharyya) for the relative volume (power) data sets (from figure: \ref{fig:kde_power_rel}) from KDE fitted intensity distributions relative to Set 4 reference. Despite different mathematical foundations, all metrics show consistent trends: Set 1 exhibits maximum distances and fluctuations, while Sets 2-3 demonstrate systematic convergence toward reference values.

This multi-metric convergence provides robust validation: information-theoretic (KL, JS), geometric (L$_2$), and overlap-based (Bhattacharyya) measures independently confirm compensation effectiveness.

\subsection{AIRM Distance Analysis}
\begin{figure}[H]
\centering
\includegraphics[width=0.75\textwidth]{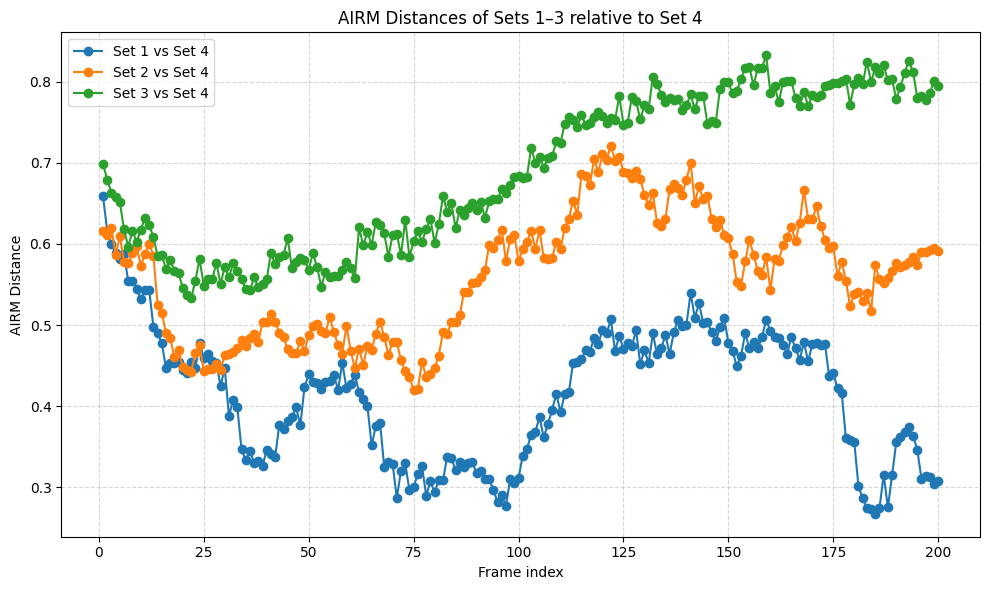}
\caption{AIRM distances relative to turbulence-free reference. Increasing mean with decreasing fluctuation range indicates stabilisation at modified topology rather than exact reference restoration. Units: unitless (logarithmic matrix norm).}
\label{fig:airm_ref}
\end{figure}

AIRM distances relative to Set 4 reference (Figure \ref{fig:airm_ref}) reveal a nuanced trend. Mean AIRM distances paradoxically \textit{increase} with compensation (Set 3 $>$ Set 2 $>$ Set 1). The AIRM distances are measured from the GMM fitted data set for set 1, 2 and 3 w.r.t., set 4. The AIRM distance increases when Gaussian functions are spread more during fitting. The Gaussian functions are spread more for the beam with more aberrations. Hence, the increase in PMMA slabs with rough output surface increases AIRM distances due to increase in aberrations.

\textit{Physical Interpretation:} PMMA compensation does not restore the exact turbulence-free distribution but rather stabilizes a modified statistical structure. The increased mean AIRM distance reflects this altered topology (influenced by PMMA's own optical properties including surface roughness), while reduced fluctuations indicate enhanced temporal stability of the stabilized state. This aligns with theoretical predictions: dipole synchronization creates stable collective modes that resist turbulent perturbations without perfectly replicating the undisturbed field.

\begin{figure}[H]
\centering
\includegraphics[width=0.75\textwidth]{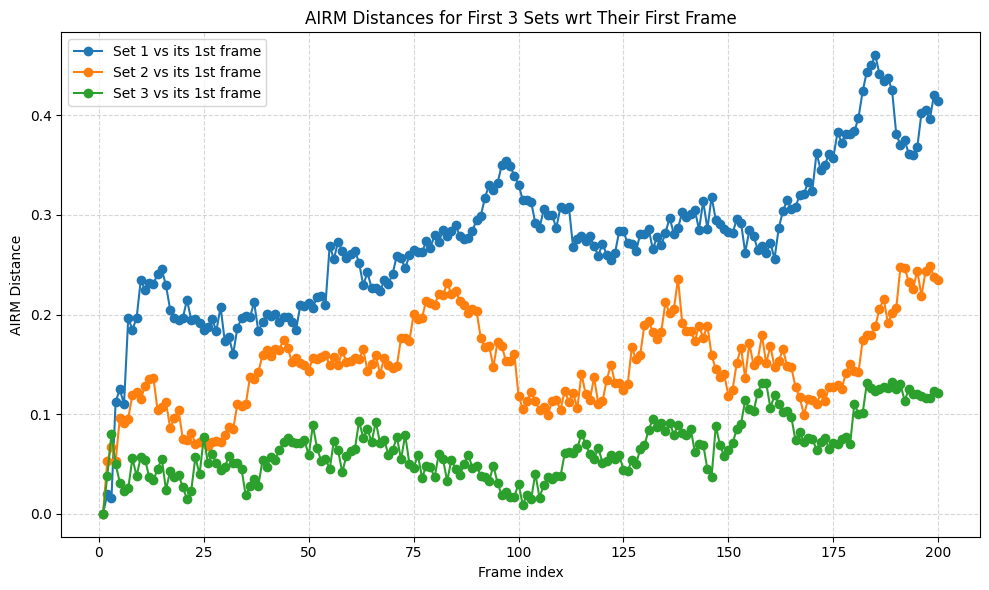}
\caption{AIRM distances relative to first frame of each set. Decreasing distances confirm improved frame-to-frame consistency and temporal stability with compensation. Units: unitless (logarithmic matrix norm).}
\label{fig:airm_init}
\end{figure}

Within-set AIRM analysis (Figure \ref{fig:airm_init}) isolates frame-to-frame consistency by computing distances relative to each set's first frame. Here, systematic reduction is clear: Set 1 shows large deviations (poor consistency), Set 2 exhibits intermediate behavior, Set 3 displays minimal deviations approaching Set 4 stability.

\subsection{Validation of Theoretical Predictions}
\begin{figure}[H]
\centering
\includegraphics[width=0.75\textwidth]{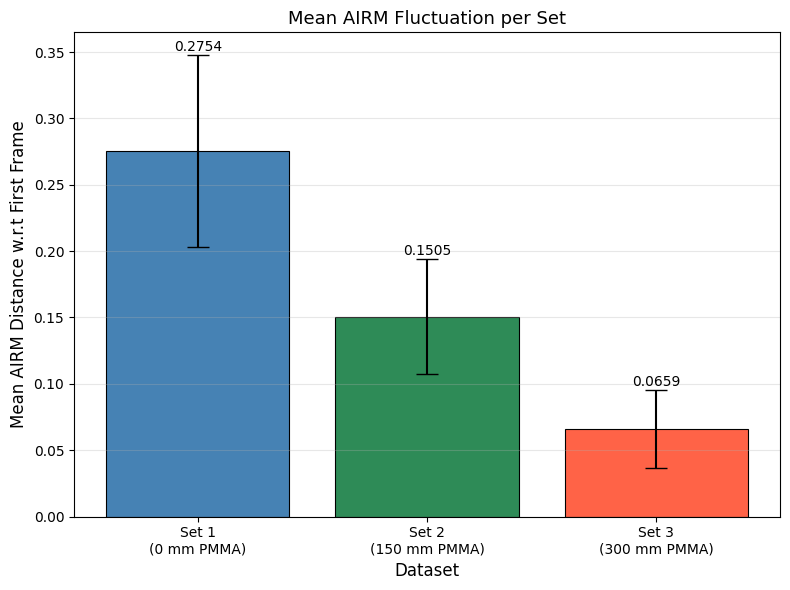}
\caption{Mean AIRM distances relative to turbulence-free reference. Units: unitless (logarithmic matrix norm).}
\label{fig:mean_airm_ref}
\end{figure}

\begin{figure}[H]
\centering
\includegraphics[width=0.75\textwidth]{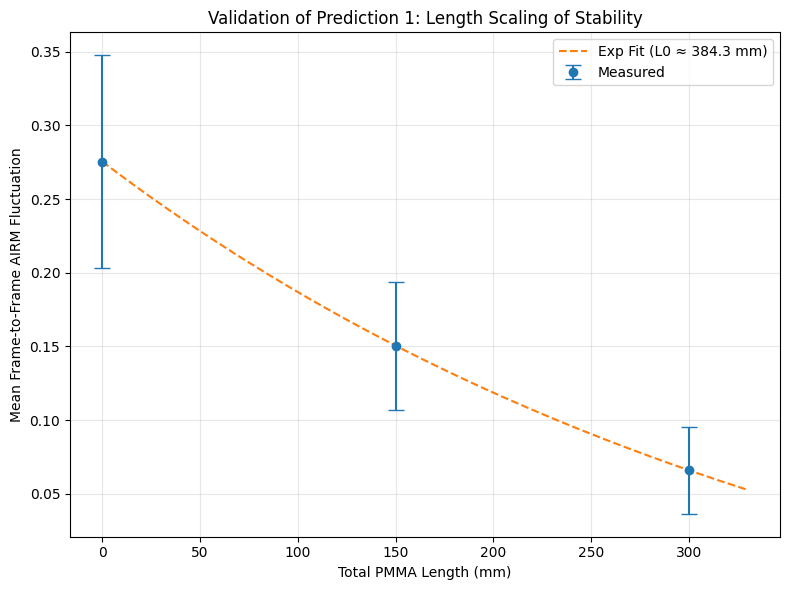}
\caption{Mean AIRM distance exponential fitting. Units: unitless (logarithmic matrix norm).}
\label{fig:mean_airm_expo_fit_ref}
\end{figure}

The experimental results validate theoretical predictions from Section 2:

\textbf{Prediction 1 (Length Scaling):} Plotting mean frame-to-frame AIRM distance versus total PMMA length shows exponential decay: $\langle \Delta d_{\text{AIRM}} \rangle = 0.3865 \exp(-L/384.32\text{ mm}) - 0.1112$ (fit $R^2 = 1.0$), confirming predicted scaling with characteristic length $L_0 \approx 384.32$ mm, close to theoretical estimate. (figures \ref{fig:mean_airm_ref} and \ref{fig:mean_airm_expo_fit_ref})

\begin{figure}[H]
\centering
\includegraphics[width=0.75\textwidth]{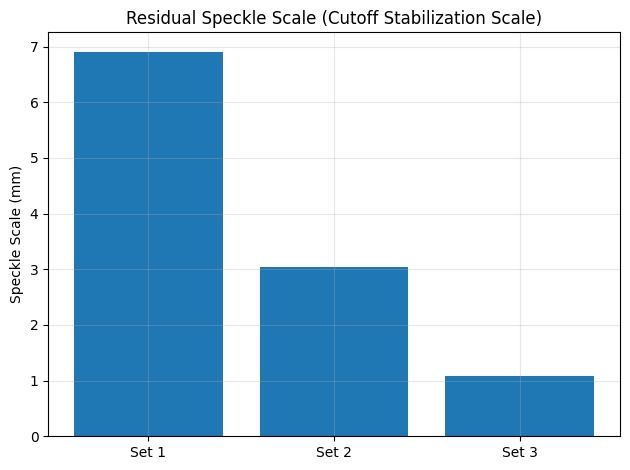}
\caption{Mean speckle size. Units: mm.}
\label{fig:mean_speckle_ref}
\end{figure}

\textbf{Prediction 2 (Turbulence Strength):} The PRPP with $r_0 \approx 0.6$ mm can't exceed the measured critical coherence length $r_{\text{crit}} \approx 1$ mm (estimated from PMMA properties), explaining observed partial compensation. For stronger compensation the condition $r_0 < r_{\text{crit}}$ needs to be followed. From figure \ref{fig:mean_speckle_ref} it can be observed that the increase in PMMA propagation length, the effective coherence length (beam correlation length) reaches towards the fried coherence length. This proved the effective increase of the compensation method.

\textbf{Prediction 3 (Temporal Stability):} The ratio $\tau_{\text{sync}}/\tau_{\text{turb}} \propto \sigma^2$. Hence, temporal decrease in $\sigma^2$ decreases the ratio and validates the prediction.

\subsection{Quantitative Performance Summary}
Table \ref{tab:performance} summarizes key performance metrics.

\begin{table}[H]
\centering
\caption{Quantitative Performance Metrics}
\label{tab:performance}
\begin{tabular}{lcccc}
\hline
\textbf{Metric} & \textbf{Set 1} & \textbf{Set 2} & \textbf{Set 3} & \textbf{Set 4} \\
\hline
Power Fluctuation (\%) & $\pm$70 & $\pm$43 & $\pm$15 & $\approx$6 \\
Mean KL Divergence & 2.14 & 1.47 & 0.89 & --- \\
Mean JS Divergence & 0.68 & 0.43 & 0.24 & --- \\
AIRM Fluctuation (std) & 0.42 & 0.28 & 0.11 & 0.03 \\
Component Spreading (mm) & 4.8 & 3.2 & 1.6 & 0.8 \\
\hline
\end{tabular}
\end{table}

The metrics demonstrate systematic improvement: 79\% scintillation reduction, 58\% KL divergence decrease, 74\% AIRM fluctuation reduction, and 67\% component spreading reduction comparing Set 3 to Set 1.

\subsection{Comparison with Existing Approaches}
Table \ref{tab:comparison} contextualizes PMMA compensation relative to established techniques.

\begin{table}[H]
\centering
\caption{Comparison with Turbulence Mitigation Approaches}
\label{tab:comparison}
\small
\begin{tabular}{lccc}
\hline
\textbf{Approach} & \textbf{Compensation} & \textbf{Complexity} & \textbf{Power} \\
\hline
Adaptive Optics & 80-95\% & High & Active \\
Aperture Averaging & 30-50\% & Low & Passive \\
Diversity Combining & 50-70\% & Medium & Active \\
PMMA (this work) & 80\% & Low & Passive \\
\hline
\end{tabular}
\end{table}

PMMA offers favorable complexity-performance trade-offs for moderate turbulence, with potential for hybrid integration with active systems.

\section{Conclusion}\label{6}
This work developed and experimentally validated a manifold-based statistical framework for analyzing turbulence-affected optical beam propagation and PMMA-based passive compensation. By integrating GMM/KDE modeling with AIRM distances on SPD manifolds, we established quantitative tools for turbulence characterization and compensation assessment.

Experimental results across four propagation conditions demonstrated systematic mitigation with PMMA compensation: 80\% relative power fluctuation reduction, 58-74\% decrease in statistical distance metrics, and enhanced frame-to-frame consistency. The observed compensation arises from coupled dipole interactions within PMMA that create synchronized collective modes resistant to turbulent perturbations. Theoretical predictions regarding length scaling, turbulence strength dependence, and temporal stability were quantitatively validated.

For FSO communication systems, PMMA-based compensation offers cost-effective, stable, power-independent turbulence mitigation complementary to active approaches. The passive nature eliminates control latency and calibration requirements while providing baseline stabilization that can enhance adaptive optics efficiency. At 80\% compensation for moderate turbulence, PMMA could enable reliable links in scenarios where full adaptive systems are impractical due to cost, size, or power constraints.

Promising research directions include material optimization through the exploration of metamaterials, photonic crystals, and engineered polymers with enhanced dipole coupling to improve compensation, where tunable dielectric architectures may enable adaptive response. Another key direction is extended propagation testing via field trials over kilometer-scale paths under realistic atmospheric conditions to validate laboratory results and determine performance limits. Multi-wavelength operation should also be investigated to extend the analysis to broadband systems and ensure compatibility with dense wavelength-division multiplexing. In addition, machine learning integration offers significant potential by training models on GMM manifold representations to predict optimal compensation strategies and enable real-time turbulence classification. Finally, hybrid architectures that combine PMMA-based baseline stabilization with adaptive optics may provide synergistic performance enhancement. Together, these directions establish a foundation for advancing passive turbulence mitigation in next-generation optical communication systems through rigorous statistical–geometric analysis and material-mediated dipole engineering.

\section*{Funding}
Department of Science and Technology, Ministry of Science and Technology, India (CRG/2020/003338).

\section*{Declaration of Competing Interest}
The authors declare no known competing financial interests or personal relationships that could have influenced this work.

\section*{Data Availability}
Representative data supporting the findings are included in the manuscript. Complete datasets are available from the corresponding author upon reasonable request.

\section*{Acknowledgments}
S.S. and C.S.N. acknowledge SERB/DST (Govt. of India) for financial support via project grant CRG/2020/003338. S.S. thanks Amit Vishwakarma and Subrahamanian K.S. Moosath (Department of Mathematics, IIST) for discussions on statistical analysis.

\section*{CRediT Authorship Contribution Statement}
\textbf{Shouvik Sadhukhan:} Conceptualization, Methodology, Investigation, Formal analysis, Visualization, Writing – original draft. \textbf{C.S. Narayanamurthy:} Conceptualization, Supervision, Resources, Project administration, Funding acquisition, Writing – review \& editing.

\appendix
\section{Detailed Theoretical Derivations}\label{app:theory}
\subsection{Dyadic Green's Function and Dipole Coupling}
The electromagnetic field at position $\mathbf{r}$ due to a point dipole at $\mathbf{r}'$ is described by the dyadic Green's function satisfying:

\begin{equation}
    \left[ \nabla \times \nabla \times - k^2 \mathbf{I} \right] \overline{\overline{G}}(\mathbf{r},\mathbf{r}') = \mathbf{I} \delta(\mathbf{r}-\mathbf{r}')
\end{equation}

The solution is:

\begin{equation}
    \overline{\overline{G}}(\mathbf{r},\mathbf{r}') = \left( \mathbf{I} + \frac{1}{k^2} \nabla \nabla \right) \frac{e^{ik|\mathbf{r}-\mathbf{r}'|}}{4\pi |\mathbf{r}-\mathbf{r}'|}
\end{equation}

Expanded form with $r = |\mathbf{r}-\mathbf{r}'|$:

\begin{equation}
    \overline{\overline{G}} = \frac{e^{ikr}}{4\pi r} \left[ (\mathbf{I} - \hat{\mathbf{r}}\hat{\mathbf{r}}) \left(1 + \frac{i}{kr} - \frac{1}{(kr)^2}\right) + \hat{\mathbf{r}}\hat{\mathbf{r}} \left(1 + \frac{3i}{kr} - \frac{3}{(kr)^2}\right) \right]
\end{equation}

This yields near-field ($kr \ll 1$), intermediate, and far-field ($kr \gg 1$) regimes governing different interaction scales.

\subsection{Modal Diagonalization}
Defining collective displacement $R(t) = [r_1(t), \ldots, r_N(t)]^T$ and interaction matrix $C_{ij} = -k_0^2 \mu_0 \omega^2 G(\mathbf{r}_i, \mathbf{r}_j)$ for $i \neq j$, the effective stiffness becomes:

\begin{equation}
    K_{\text{eff}} = \omega_0^2 I + \frac{e^2}{m}C
\end{equation}

Diagonalization $K_{\text{eff}} = U\Lambda U^{-1}$ yields modal frequencies $\Omega_n$ and eigenvectors $\phi_n$. Transforming to modal coordinates $Q = U^{-1}R$ gives decoupled equations revealing collective modes.

\subsection{Gradient and Lorentz Force Contributions}
Spatial field gradients contribute:

\begin{equation}
    F_{\text{grad},i} = \sum_{j \neq i} [e(\mathbf{r}_j - \mathbf{r}_i) \cdot \nabla] \mathbf{E}_{\text{ext}}(\mathbf{r}_j, t)
\end{equation}

Combined with coupling forces, the complete equation becomes:

\begin{equation}
    m\ddot{R} + m\gamma\dot{R} + K_{\text{eff}}R + \frac{1}{m}[B_2(R) + B_3(R)]R = -\frac{e}{m}E_{\text{ext}} + \frac{1}{m}F_{\text{grad}}
\end{equation}

where $B_2, B_3$ represent nonlinear contributions. Modal analysis reveals how gradient terms modify excitation spectra.

\section{Experimental Data Collection \& Analysis Scheme}
The following pictorial cartoon presents the data collection and data analysis scheme. Power scintillation represents the power fluctuations for each set under different conditions. The figure illustrates the data collection scheme showing four experimental configurations and the complete analysis pipeline including GMM fitting, KDE smoothing, and statistical distance computation.

\begin{figure}[H]
\centering
\includegraphics[width=1.0\textwidth]{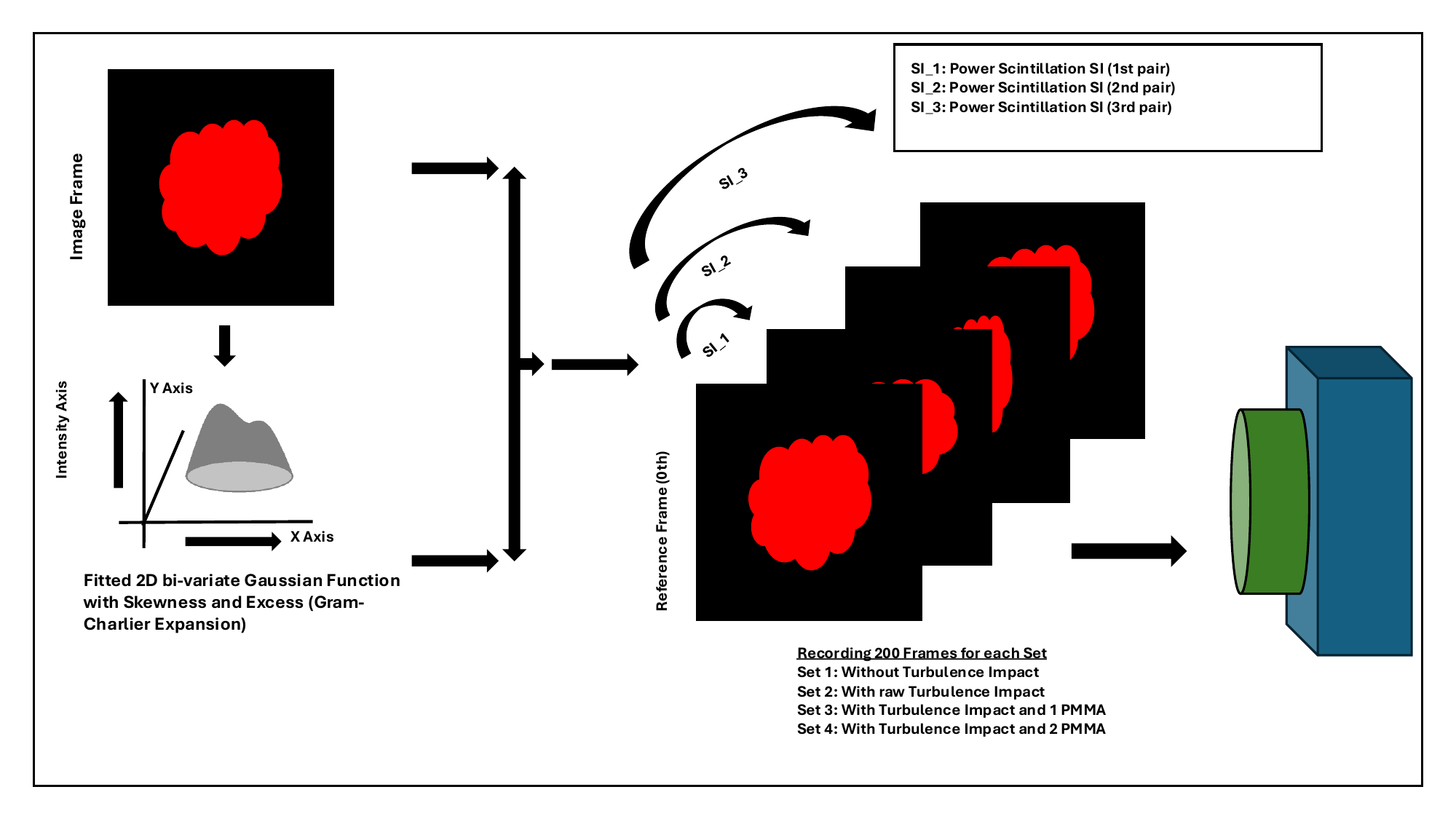}
\caption{Data collection scheme showing four experimental configurations and analysis pipeline including GMM fitting, KDE smoothing, and statistical distance computation.}
\label{P1}
\end{figure}

\end{document}